\documentclass[english]{scrartcl}
\usepackage[utf8]{inputenc}
\usepackage{geometry}
\usepackage{pdflscape}
\usepackage{amsmath,amssymb,amsfonts}
\usepackage[authoryear]{natbib}
\usepackage{booktabs}
\usepackage{diagbox}
\usepackage{multirow}
\usepackage{makecell}
\usepackage{longtable}
\usepackage{subcaption}
\usepackage{bbm}
\usepackage{graphicx}
\usepackage{xcolor}
\usepackage[unicode=true,pdfusetitle,bookmarks=true,bookmarksnumbered=true,bookmarksopen=true,bookmarksopenlevel=1,breaklinks=true,pdfborder={0 0 1},backref=false,colorlinks=true]
 {hyperref}
\hypersetup{
 linkcolor=blue, citecolor=blue, urlcolor=blue, filecolor=blue,pdfpagelayout=OneColumn, pdfnewwindow=true,pdfstartview=XYZ, plainpages=false}
\usepackage{authblk}
\usepackage{backref}
\usepackage{enumitem}

\setlist{leftmargin=*, topsep=0.5em, parsep=0pt, itemsep=1em, labelindent=0pt, align=left}

\definecolor{blue}{HTML}{1F77B4}
\definecolor{orange}{HTML}{FF7F0E}
\definecolor{green}{HTML}{2CA02C}
\definecolor{red}{HTML}{D62728}
\definecolor{purple}{HTML}{9467BD}
\definecolor{brown}{HTML}{8C564B}
\definecolor{pink}{HTML}{E377C2}
\definecolor{grey}{HTML}{7F7F7F}
\definecolor{yellow}{HTML}{BCBD22}
\definecolor{cyan}{HTML}{17BECF}
\definecolor{turquoise}{HTML}{3FE0D0}



\usepackage{ifpdf} 
\ifpdf 
 \IfFileExists{lmodern.sty}{\usepackage{lmodern}}{}
\fi 

\let\myTOC\tableofcontents
\renewcommand\tableofcontents{%
  \pdfbookmark[1]{\contentsname}{}
  \myTOC
}

\renewcommand*{\backref}[1]{}
\renewcommand*{\backrefalt}[4]{%
   \ifcase #1 
    \or 
      (Cited on page~#2)%
   \else
      (Cited on pages~#2)
    \fi} 

\definecolor{blue}{HTML}{1F77B4}
\definecolor{orange}{HTML}{FF7F0E}
\definecolor{green}{HTML}{2CA02C}
\definecolor{red}{HTML}{D62728}
\definecolor{purple}{HTML}{9467BD}
\definecolor{brown}{HTML}{8C564B}
\definecolor{pink}{HTML}{E377C2}
\definecolor{grey}{HTML}{7F7F7F}
\definecolor{yellow}{HTML}{BCBD22}
\definecolor{cyan}{HTML}{17BECF}
\definecolor{turquoise}{HTML}{3FE0D0}

\begin{document}
\title{The dynamics of innovation diffusion:\\ A survey of Bass-type models}
	
\author[1]{Nicolas Langren\'e}	
\author[2]{Rui Liu\thanks{{Corresponding author, rl2625@ic.ac.uk}}}
\author[3]{Xiangqin Wu}
\author[1]{Tianhao Zhi}

\affil[1]{\normalsize Guangdong Provincial/Zhuhai Key Laboratory of Interdisciplinary Research and Application for Data Science, Beijing Normal-Hong Kong Baptist University, China} 
\affil[2]{\normalsize Imperial College London, UK} 
\affil[3]{\normalsize University of Southampton, UK}

\maketitle
	
\begin{abstract}
This paper synthesises the existing research on the dynamics of innovation diffusion, with a focus on Bass-type models and their extensions. The theoretical foundation of innovation diffusion proposed by \citet{rogers1962innovation} and the seminal work of \citet{bass1969new} serve as a starting point for the analysis. We identify and examine various generalizations and stochastic extensions of the Bass model, including counting processes, diffusion processes, and uncertain processes, as well as parameter estimation techniques, from classical statistical techniques to more advanced techniques such as Bayesian filtering and metaheuristic optimisation. We finally explore alternative models of innovation diffusion, with a particular focus on agent-based models. This overview of the evolution of Bass-type models illustrates the progress made in innovation diffusion research over the past decades.

\end{abstract}

\newpage


\section{Introduction}
\label{sec:introduction}

Why do some good ideas fail to become good products?

This question may equally bother scientists and inventors – the knowledge creators, as well as entrepreneurs – the knowledge monetisers. Serendipity is prevalent in both groundbreaking scientific discoveries and entrepreneurial successes. While it takes painstaking effort for scientists and inventors to make fundamental progress in basic research and to conceptualise novel creations, the question of whether these discoveries and inventions will be successfully adopted and disseminated as new products or services is a distinct dimension of complexity and uncertainty.

\subsection{Theory of innovation diffusion}
\label{subsec:innovation_diffusion}

Such a quest has inspired decades of research on innovation diffusion, as best depicted in the book \textit{``Diffusion of Innovations’’} by \citet{rogers1962innovation}, who crystallised the state of the art in innovation diffusion research at the time the book was published. The book, and its latest revision \citet{rogers2003innovation}, illustrates plentiful examples of the successes and failures of diffusion from ideas to products. One example that Rogers examined is the non-adoption of the Dvorak keyboard: the widely used \textit{QWERTY} keyboard, from which this article was typed, was first invented by C.L. Sholes as early as 1878, coming along with the invention of the physical typewriter. The letters of the \textit{QWERTY} keyboard have been arranged in such a way that it slowed the typists in order to prevent paper jams - a necessary adaptation to cater for the physical limits of typewriters at the time. Later, the \textit{AOEUIDHTN} keyboard was invented by A. Dvorak at the University of Washington in 1932, offering a much better user experience. Despite its convenience and efficiency, the Dvorak keyboard has never diffused successfully to date, even as personal computers (PCs), which are entirely free of paper jamming issues, have long replaced traditional typewriters.

The diffusion of PCs follows an even more tortuous journey, yet it was luckily celebrated for its eventual success. As Rogers illustrated, the first PC was developed at Xerox PARC (Palo Alto Research Centre) and was named \textit{Alto}. Yet Xerox never realised Alto's full commercial potential, and the technology was transferred to Apple, which transformed it into the success of the Apple Macintosh and the widespread adoption of PCs today. The diffusion of laptops shares a similar story, as Rogers pointed out in another example: it was first envisioned by Toshiba's engineer Tetsuya Mizoguchi, and only became commercially successful and widely adopted thanks to his perseverance despite the strong objection from Toshiba's senior management at the time when its development was in its infancy \citep{rogers2003innovation}.

The diffusion of computing technologies, along with other examples illustrated in the book, has led to the formulation of Rogers' theory of diffusion of innovations. Rogers identifies heterogeneity among adopters in terms of their different stages of adoption: innovators are the most crucial initial adopters, followed by early adopters, early majority, late majority, and laggards. The success of innovation depends not only on the novelty of innovation itself, but also on other factors such as communication channels, social and cultural norms, and critical mass. The leader-follower roles of innovation adopters and their determining factors form the basis of the S-shaped curve that describes the relationship between time and proportion of adoption, which is at the core of the dynamics of innovation diffusion. Rogers' theory offers profound insights into the processes and dynamics of innovation diffusion, as well as a methodological foundation for R\&D risk management and the economic appraisal of innovations.

\subsection{The Bass model}
\label{subsec:bass_model}

While \citet{rogers1962innovation} offers a descriptive theory of innovation diffusion, \citet{bass1969new} translates this theory into a mathematical expression:
\begin{eqnarray}
    \frac{dF(t)}{dt} &=& [1 - F(t)][p + qF(t)],
    \label{eq:dFt}
    \\
    F(t) &=& \frac{1-e^{-(p+q)t}}{1+\frac{q}{p}e^{-(p+q)t}},
    \label{eq:Ft}
\end{eqnarray}
\noindent where
\begin{itemize}
\item $F(t)$ represents the fraction of adopters in a population at time $t$;
\item $p$ is the innovation parameter, capturing the strength of initial adoptions;
\item $q$ is the imitation parameter, capturing the strength of followers at a later stage.
\end{itemize}

The Bass model provides a quantitative realisation of Rogers' theory that reduces the complex nature of innovation diffusion to two crucial behavioural parameters embedded in equation~\eqref{eq:dFt}: one describing the behaviour of early adopters ($p$) and the other describing the behaviour of late comers ($q$). It suggests the following interpretation of the nature of R\&D risks: an innovation can fail simply due to a lack of early adopters, latecomers, or both. This is corroborated by empirical research such as \citet{kohli1999incubation}, which sets an important stage for future empirical studies of innovation diffusion.

\begin{figure}[h]
    \centering
    \includegraphics[width=0.74\paperwidth]{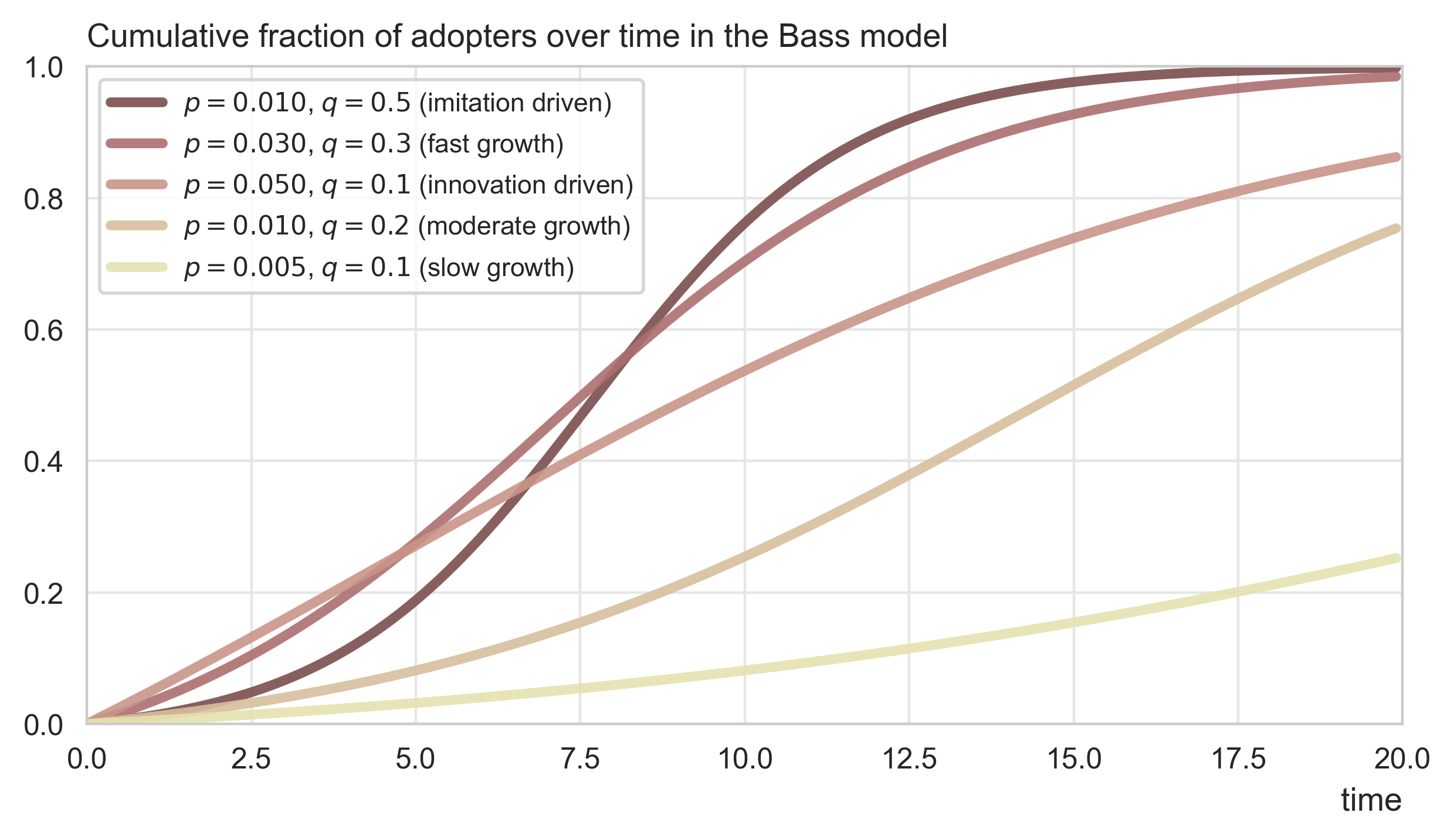}
    \caption{Evolution of adopter fraction under varying innovation $p$ and imitation $q$.}
    \label{fig:bass_model_curves}
\end{figure}

Figure~\ref{fig:bass_model_curves} illustrates the temporal trajectory of new product adoption rates under varying values of $p$ and $q$. A relatively high $p$ characterises a diffusion process primarily propelled by innovation, whereas a dominant $q$ indicates that adoption is largely driven by imitative behaviour.

\subsection{Aims of the paper}
\label{subsec:aims}

With this motivation, this paper conducts a meta-study of Bass-type models, aiming to synthesise their development since their origin in \citet{bass1969new}, with a particular emphasis on their structural extensions and parameter estimation techniques. We begin with multi-factor extensions, such as the inclusion of pricing effects and multi-generational products, before moving on to the various stochastic and uncertain extensions of the model. We then review the parameter estimation techniques proposed in the literature for the model. We end up with a discussion of \textit{bottom-up} alternatives to the aforementioned \textit{top-down} Bass-type models, with a particular emphasis on agent-based approaches to modelling innovation diffusion.

Our survey has important theoretical and practical implications; it not only formalises Roger's theory of innovation diffusion, that delves deeper into the nature of innovation diffusion, with a comprehensive quantitative framework, but also sheds light on the valuation of intellectual properties and the risk management of R\&D projects. Similarly to how classical asset pricing models acknowledge the unpredictable nature of asset prices and formalise the rational decision making process of agents through their risk profiles, we can understand the value of intellectual properties by examining their underlying stochastic and complex dynamics, with respect to their quantitative features that are investigated in this survey, to the end of helping entrepreneurs and risk managers to make informed decisions over their innovative ventures.

The organisation of this paper is as follows: Section~\ref{sec:bass_model} reviews the Bass model and its several deterministic and stochastic extensions. Section~\ref{sec:bass_estimation} reviews the existing methods that have been used in the literature to estimate the parameters of the Bass model. Finally, Section~\ref{sec:others} explores some other promising innovation diffusion models and approaches, in particular agent-based models.

\section{The Bass model}
\label{sec:bass_model}

Following Rogers' theory, \citet{bass1969new} developed a differential equation to describe how new products are adopted and accepted by people once they enter the market. In this theory, adopters of new products are divided into two categories: innovators and imitators. Initially, innovators communicate with each other, develop, and promote the new product, while imitators are influenced by the decisions of other members within the social system. Over time, as more and more people adopt the new product, their influence on other members also increases. Eventually, the new product completely enters the market and is adopted by almost everyone. Bass described this process with the following equation:
\begin{equation}
    \label{eq:bass}
    \frac{dF(t)}{dt} = (1 - F(t))(p + qF(t)),\quad t\in[0,T]
\end{equation}

where $F(t)$ is the cumulative likelihood of purchasing (the new product) between time $0$ to time $T$, and $p$ and $q$ represent the coefficient of innovation and coefficient of imitation, respectively. The corresponding cumulative density function is
\begin{equation}
    \label{eq:bass_cum}
    F(t) = \frac{1-e^{-(p+q)t}}{1+\frac{q}{p}e^{-(p+q)t}}
\end{equation}

The beauty of the Bass model lies in its convincing theoretical foundations, which broadly align with the real-world process of new product diffusion. It typically starts with a subset of innovators using the product, followed by influencing other members of the social system through advertising, word-of-mouth, and other means, eventually leading to full market penetration. Regression analyses of the Bass model also demonstrate its alignment with the observed patterns of many products in the real world \citep{kohli1999incubation}.

To transition from the probability of adoption to actual market demand, the model introduces the total market potential, or the ultimate number of adopters, denoted as $m$. By scaling the cumulative likelihood $F(t)$ by the market capacity $m$, the cumulative number of adopters $S(t)$ at time $t$, a.k.a. the cumulative sales, can be expressed as:
\begin{equation}
    \label{eq:St}
    S(t) = m \cdot F(t) = m \frac{1 - e^{-(p+q)t}}{1 + \frac{q}{p} e^{-(p+q)t}}
\end{equation}

The instantaneous sales volume, denoted as $s(t)$, is defined as the rate of change in the cumulative number of adopters, which corresponds to the first derivative of $S(t)$ with respect to time:
\begin{equation}
    \label{eq:st}
    s(t) = \frac{dS(t)}{dt} = m \frac{(p+q)^2}{p} \frac{e^{-(p+q)t}}{\left(1 + \frac{q}{p} e^{-(p+q)t}\right)^2}
\end{equation}

\begin{figure}[h]
    \centering
    \includegraphics[width=0.74\paperwidth]{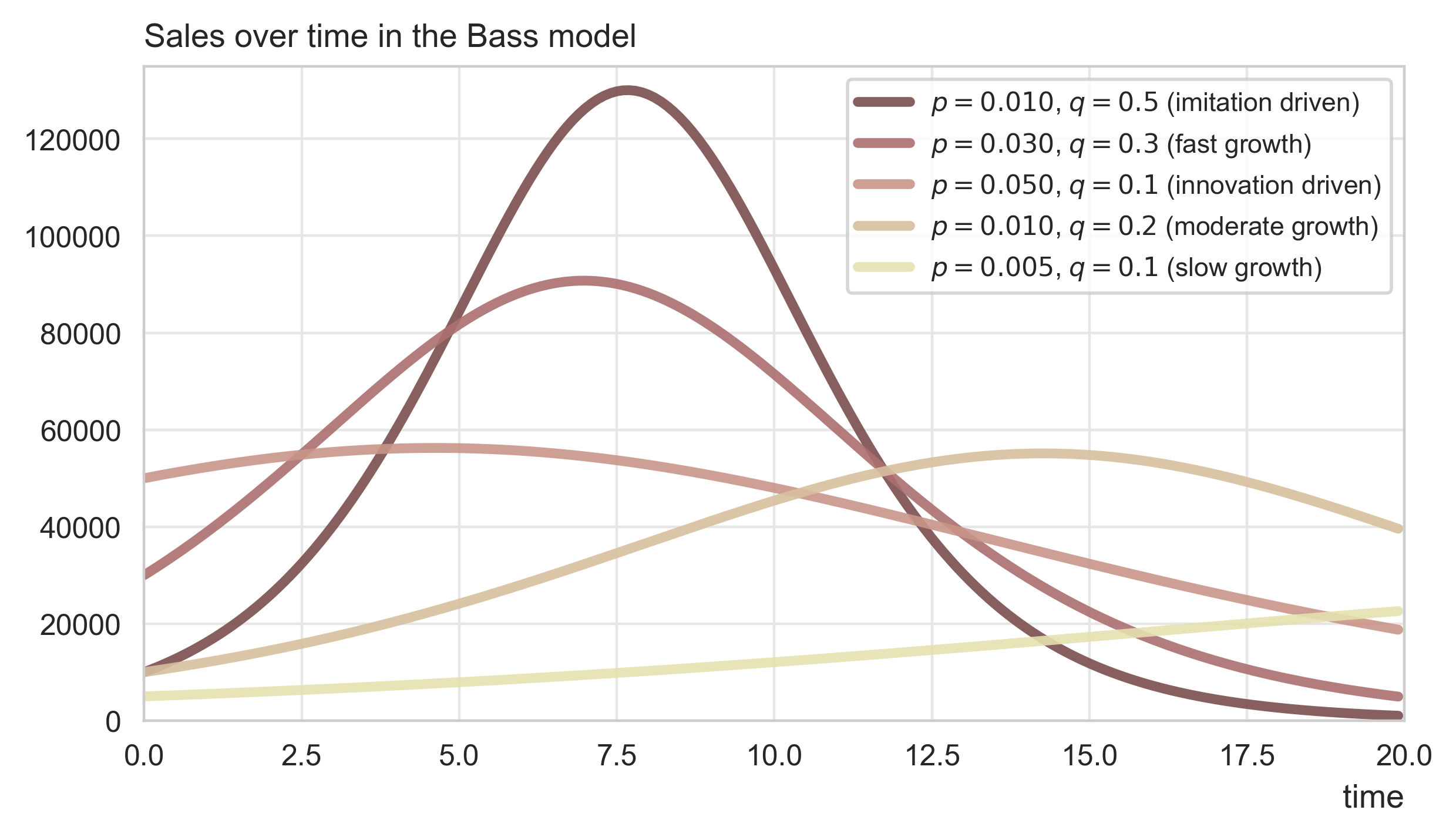}
    \caption{Evolution of sales under varying innovation $p$ and imitation $q$, with a market potential of $m=1$ million.}
    \label{fig:bass_model_sales}
\end{figure}

As shown on Figure~\ref{fig:bass_model_sales}, this sales function $s(t)$ typically exhibits a bell-shaped curve, capturing the acceleration and subsequent deceleration of market growth. According to the model, the time of peak sales is equal to $t^*=(\log(q)-\log(p))/(p+q)$.

When accounting for the market capacity, the Bass model provides a useful framework for forecasting both the timing and the magnitude of peak sales, with application to strategic capacity planning and inventory management in various industries.

\subsection{Bass model variations}
\label{subsec:bass_variations}

In the past few decades, various extensions have been made to the Bass model to make it applicable to more complex real-world scenarios. For instance, \citet{bass1994bass} introduced new variables into the original model to consider the impact of pricing on the diffusion of new products:
\begin{equation}
    \label{eq:bass_pricing}
    \frac{f(t)}{1-F(t)}=(p+{q}F(t))x(t),
\end{equation}

where $f(t)=F'(t)$, and $x(t)$ is a function of percentage change in price and other variables.

Another example is the model proposed by \citet{norton1987diffusion}, which takes into account the multi-generational nature of some products and the existence of repeat purchase behaviour. For instance, the cumulative sales formulas for three generations are as follows:
\begin{equation}
    \label{eq:bass_multigenerational}
    \begin{aligned}
        &S_{1}(t) = F(t_1)b_1(1-F(t_2))\\
        &S_{2}(t) = F(t_2)(b_2+F(t_1)b_1)(1-F(t_3))\\
        &S_{3}(t) = F(t_3)(b_3+F(t_2))(b_2+F(t_1)b_1),
    \end{aligned}
\end{equation}

with $b_i$ = $a_i m_i$, where $m_i$ and $a_i$ are the incremental number of ultimate adopters of the $i^{\mathrm{th}}$~generation product, and the average (continuous) repeat buying rate among adopters of the $i^{\mathrm{th}}$~generation product, respectively.

Building on this foundation, subsequent research sought to decompose the underlying components of sales more precisely. \citet{bass2001diffusion} pointed out that the Norton-Bass model aggregates first-time and repeat purchases, making it difficult to track the evolution of the installed base. They proposed a generalized model that explicitly separates adoption sales (first-time purchases) from repeat sales, providing a clearer view of the internal demand structure. Furthermore, considering the competitive landscape, \citet{altinkemer2008multigeneration} extended the framework to a duopoly setting. Their model not only captures the inter-generational cannibalisation effects but also integrates brand competition (e.g., between Sony and Microsoft), demonstrating how the market potential of one firm is dynamically influenced by the presence of a rival.

Moreover, \citet{gatignon1989modeling} also develops models suitable for cross-national expansion, in which the market potential is estimated for each country; the coefficients of innovation and imitation are functions of national characteristics:

\begin{equation}
    \label{eq:bass_crossnational}
    p_i = \beta_{p,i,0} + \sum_k{\beta_{p,i,k}Z_{i,k}} +e_{p,i},
\end{equation}

where $Z_{i,k}$ represents a cultural variable, $\beta_{p,i,k}$, for
$i = 0, 1, \ldots$, are estimated coefficients, and $e$ is a
disturbance term. The model~\eqref{eq:bass_crossnational} suggests that the characteristics of a national culture capture a nation's intensity to innovate new products.

\subsection{Stochastic Bass model}
\label{subsec:stochastic_bass}

The traditional Bass model serves as a deterministic model, implying certain difficulties in its application to real markets due to the inherent stochastic nature of real-world market dynamics. Indeed, when modelling in uncertain markets, adopting a stochastic approach often proves effective. For example, the classical Black-Scholes model for option pricing \citep{black1973pricing} and its numerous extensions and generalizations are built upon stochastic processes. Models based on stochastic processes not only allow for the consideration of markets' inherent volatility, but also provide a natural approach for the simulation and testing of the model's stability and robustness.

Currently, the stochasticisation of the Bass model has not received significant attention, nor has there been a systematic review of this stream of research. Based on our research, current studies on stochastic Bass models primarily adopt the following approaches to construct models:

\begin{enumerate}
    \item \textbf{Integrating the deterministic Bass solution into stochastic jump processes.} Since the traditional Bass model provides a closed-form analytical solution for cumulative adoption, one prevalent approach is to treat this solution as a time-varying parameter within other stochastic frameworks (e.g., nonhomogeneous Poisson processes), rather than introducing stochasticity into the differential equation itself. Detailed discussions on these hybrid frameworks can be found in Subsection~\ref{subsubsec:integration_jump_process}.
    \item \textbf{Modelling the diffusion rate as a stochastic diffusion process.} In the classical model, the diffusion rate is a deterministic function of the current state. By augmenting the governing equation with a stochastic term—often representing random fluctuations in the coefficients of innovation or imitation—researchers can better capture the volatility of market dynamics. Besides, to account for empirical fluctuations in observed sales data, exogenous noise terms can be integrated into the Bass model. This formulation also enables the application of Ito’s calculus for analytical derivations, as elaborated in Subsection~\ref{subsubsec:modelling_diffusion_process}.
    \item \textbf{Constructing models within non-probabilistic uncertainty frameworks.} Beyond traditional probability theory, recent studies have extended the Bass model using uncertainty theory to handle ``human uncertainty'' (e.g., belief degrees). These Uncertain Bass models are particularly effective when historical data is scarce or when market dynamics are driven by subjective factors rather than frequentist randomness. The mathematical characteristics of such models are described in Subsection~\ref{subsubsec:uncertain_theory}.
\end{enumerate}

\subsubsection{Integration with stochastic jump processes}
\label{subsubsec:integration_jump_process}

The primary integration methodology involves coupling the deterministic Bass model with discrete-valued stochastic processes. In this framework, the adoption rate derived from the original Bass equation is utilized as a time-varying parameter (e.g., the intensity or arrival rate) within an underlying discrete-state counting process. Representative studies employing this approach include \citet{boker1987stochastic}, \citet{niu2002stochastic}, and \citet{xie2017optimal}, which are very closely related. 

\citet{boker1987stochastic} models the cumulative number of adopters $S(t)$ as a counting process, whose intensity is adapted from the Bass model. More precisely, the intensity that there will be a purchase at time $t$ is defined by:
\begin{equation}
    \label{eq:lambda_t}
    \lambda(t) = (m - S(t-))(p + q S(t-)),
\end{equation}

where $S(t-)$ is the random number of customers who bought the new product immediately before $t$ and $\lambda(t)\Delta(t)$ is the conditional probability that there is a new purchase in the small time interval $[t, t + \Delta(t)]$ given the number of buyers before time $t$. The function $\lambda(t)$ is called the intensity function of the counting process $S(t)$. The stochastic behavior of the counting process is completely determined by its intensity function. 

Similarly, \citet{niu2002stochastic} employs a pure birth process, $\{S(t)\}_{t \ge 0}$, to characterize the stochastic dynamics of consumer adoption. The model postulates a market consisting of $m$ potential adopters, each of whom is destined to adopt exactly one unit of the product. Given the inherent uncertainty regarding the timing of individual adoption decisions, $S(t)$ is defined as a random variable representing the cumulative number of adopters at time $t$, thereby capturing the aggregate progression of the innovation diffusion process within the population. 

In the model, the probability for $S$ to increase by one unit in the time interval $(t,t+h)$ in given by $\lambda(t) h +o(h)$, where
\begin{equation}
    \label{eq:lambda_mj}
    \lambda(t) := (m-S(t-))\left(\alpha+\frac{\beta}{m-1} S(t-)\right),\quad S(t-)=0,1,\ldots,m-1.
\end{equation}

In equation~\eqref{eq:lambda_mj}, $m$ represents the population size, $\alpha$ and $\beta$ represent the intrinsic adoption rate (which conceptually corresponds to the innovation parameter $p$ in the Bass model) and the induction rate (which conceptually corresponds to the imitation parameter $q$ in the Bass model) respectively.
From this definition, the birth rate $\lambda(t)$ in this process is aligned with the adoption rate in the Bass model. In fact, they demonstrate that as the parameter $m$ tends towards infinity, the solution of this process converges to that of the Bass model.

Similarly, \citet{xie2017optimal} employs a non-homogeneous Poisson process (NHPP) to characterise the cumulative adoption of new products. They assume the sales $S(t)$ follow an NHPP with sales rate $\lambda(t)$. By adopting a Bass model as the sales rate, the expected cumulative sales $E[S(t)]$ can be shown to be:
\begin{equation}
    \label{eq:ESt}
    E[S(t)]=m \frac{1-e^{-(p+q)t}}{1+\frac{q}{p}e^{-(p+q)t}}
\end{equation}
where $p$ is the innovator factor, $q$ is the imitator factor and $m$ is the potential market size. Equation~\eqref{eq:ESt} coincides with the Bass equation~\eqref{eq:St}. They called this combined model an `NHPP-Bass model'.

Furthermore, the stochastic dynamics of the Bass model can be equivalently described through the lens of self-exciting point processes, specifically the Hawkes process \citep{laurinodossantos2025hawkes}. While the aggregate adoption process may be modelled as a univariate Hawkes process if the intensity is rescaled by the remaining potential adopters $(m - S(t))$, the diffusion is perhaps most naturally captured at the individual level. In this formulation, let $\lambda_i(t)$ denote the intensity function of the $i^{\mathrm{th}}$ consumer. The Bass model corresponds to a multivariate Hawkes process where the spontaneous adoption rate corresponds to the innovation parameter $p$, and the peer influence corresponds to the imitation parameter $q$ scaled by the market size $m$. The individual intensity is defined as:
\begin{equation}
\label{eq:bass_hawkes_individual}
\lambda_{i}(t|\mathcal{H}_{t})=\mathbbm{1}_{\{N_{i}(t)<1\}}\left[p+\sum_{j\ne i}\int_{0}^{t}\frac{q}{m}dN_{j}(\tau)\right],
\end{equation}

where $\mathcal{H}_{t}$ represents the history of the process up to time $t$, $N_i(t)$ represents the counting process for individual $i$ (taking values 0 or 1), and the indicator function $\mathbbm{1}_{\{N_{i}(t)<1\}}$ ensures that the process terminates for an individual after their first purchase. This formulation explicitly links the deterministic parameters of the Bass model to the stochastic intensity of event arrivals in a point process framework.

\subsubsection{Modelling with stochastic diffusion processes}
\label{subsubsec:modelling_diffusion_process}

While jump-diffusion models are effective in capturing discrete shocks occurring on the market, they can overlook the persistent and continuous-time fluctuations that characterise the broader adoption lifecycle. To address this, the framework of stochastic diffusion models can be applied to model the time evolution of adoption as a continuous random trajectory. This transition from discrete jumps to continuous diffusions allows researchers to treat market volatility as an intrinsic, ongoing component of the governing dynamics.

\citet{skiadas1997stochastic} extends the Bass model into a Stochastic Differential Equation (SDE) framework. They add an additional term to the diffusion rate and assign it a specific meaning. The stochastic terms are attributed to the ramifications of policy influences:
\begin{equation}
    \label{eq:dft}
    dS_t = (m-S_t)\left(p+\frac{q}{m}S_t\right)dt + c\left(\frac{p}{q}+\frac{S_t}{m}\right)dW_t
\end{equation}

where $S_t$ is the cumulative number of adopters at time $t$, $m$ is the total population of potential adopters, $p$ is the coefficient of innovation, $q$ is the coefficient of imitation, $c$ is the noise parameter and $W=\{W_t\}_{t \geq 0}$ is a Brownian motion. 
The drift of this SDE corresponds to the Bass model, which postulates that the hazard rate at time $t$ is a linear function which increases with the term $(\frac{p}{q}+\frac{S_t}{m})$ describing the product entry policy. The diffusion term of this SDE is therefore a function of the product entry policy, which actually causes the random adoption pattern.

\citet{guarnera2019bass} conducts a detailed theoretical study on the type of model formulated by the more general equation:

\begin{equation}
    \label{eq:dft2}
    dS_t = \left((m-S_t)\left(p + \frac{q}{m} S_t\right) - \mu S_t\right)dt + v(S_t)dW_t,
\end{equation}

where $v(\cdot)$ is the volatility function, $W$ is a Brownian motion, and the term $\mu S_t$ takes into account the presence of adopters who stop being so. The function $v$ can be chosen such that $0\leq S_t\leq m$ for all $t\geq 0$, for example using their suggested parametric shape $v(S_t)=\sigma S_t^\beta (1-S_t/m)^\gamma$, $\sigma\geq0$, $\beta\geq1$, $\gamma\geq1$.

Instead of discussing factors influencing stochastic terms, they simply assume that the stochastic terms are smooth functions influenced by the existing number of adopters and other market factors. Through this generalised assumption, they investigate the specific properties of such models. They have proven that when all model parameters are strictly positive, a stationary distribution exists. Furthermore, they propose an inferential procedure based on the approximation of the SDE transition density using Monte Carlo diffusion bridge simulation. This procedure was applied to the SDE Bass model solutions obtained by an exact algorithm and outperforms the classical methods based on Gaussian approximation.

The preceding models treat stochasticity as endogenous to the innovation diffusion process. An alternative to this approach is to characterise the dynamics of sales as a combination of a deterministic diffusion process with an exogenous random noise process. Within this paradigm, the exogenous noise serves as the fundamental source of stochasticity.

An example is illustrated by \citet{kanniainen2010forecasting}. They posit that the sales volume $\tilde{s}_t$ is a stochastic variable, with its expected value dictated by the Bass model, while the noise conforms to an Ornstein-Uhlenbeck process:
\begin{equation}
    \label{eq:logsttilde}
     \log(\tilde{s}_t) = \log(s_t) + X_t.
\end{equation}

In equation~\eqref{eq:logsttilde}, $s_t$ is the sales volume forecast \eqref{eq:st} of the deterministic Bass model, and $X_t$ follows an OU process with mean reversion speed $\kappa$ and standard deviation $\sigma$. By applying Ito's lemma, the sales process can be written as follows:
\begin{equation}
    \label{eq:dSt}
    d\tilde{s}_t = \kappa(y_t - \log(\tilde{s}_t))s_tdt + \sigma \tilde{s}_t dW_t, S_0 > 0
\end{equation}

where $$y_t = \frac{1}{\kappa}\left(\frac{\sigma^2}{2}+\frac{d\log(s_t)}{dt}\right)+\log(s_t).$$

\citet{grasman2019forecasting} and \citet{singhal2019sde} also employ methods involving the addition of noise to stochasticise the Bass model. \citet{grasman2019forecasting}'s research aims to estimate confidence intervals for model predictions. To do so, they directly add noise to the predicted number of adopters (rather than after logarithmic transformation, as proposed in~\eqref{eq:logsttilde}):
\begin{align}
    \label{eq:Yt}
    \frac{dS_t}{dt}&=h(S_t)=(m-S_t)\left(p+\frac{q}{m}S_t\right)\\
    \tilde{S}_t &= S_t + v_t,
\end{align}

where $S_t$ represents the cumulative number of adopters and the perturbation evolves in time as $v_t$. Then, substituting and applying a Taylor expansion at $\tilde{S}_t = S_t$:
\begin{align}
    \label{eq:dYt}
    \frac{\tilde{S}_t}{dt}&= h(\tilde{S}_t) = h(S_t+v_t) \\
    &= h(S_t)+[h'(\tilde{S_t})]_{\tilde{S_t}=S_t}v_t + \frac{1}{2}[h''(\tilde{S_t})]_{\tilde{S_t}=S_t}v_t^2 + ...
\end{align}

Assuming that the perturbation $v_t$ remains small, higher order terms can be ignored. The equation takes the form of the tangent linear equation:
\begin{align}
    \label{eq:dvt}
    dv_t&= [h'(\tilde{S}_t)]_{\tilde{S}_t=S_t}v_t + \beta h(S_t)dW_t
\end{align}

where $W$ is a standard Brownian motion. Through the transformation, both $\beta$ and the variance for perturbation $v_t$ can be estimated effectively. In a similar fashion, \citet{singhal2019sde} explores generalised innovation diffusion models, which are not limited to the Bass model, while also providing empirical analysis and comparisons.

Finally, \citet{nicolosi2023approximation} extends this discourse by investigating a more sophisticated scenario that integrates multiple sources of uncertainties. Their framework first incorporates stochastic perturbations directly into the underlying dynamics of the adopter fraction. Furthermore, recognising that the precise state of adoption is often latent or imperfectly observable in practice, they introduce additional noise into the observation process (the ``observable'' state). This compounded stochasticity—arising from both system evolution and measurement error—not only enhances the model's empirical realism but also significantly increases the complexity of the associated stochastic optimal control problem.

\subsubsection{Uncertain theory approach}
\label{subsubsec:uncertain_theory}

Beyond traditional SDEs, the Uncertain Bass Model (UBM) driven by a Liu process has emerged as a framework to model epistemic uncertainty in diffusion \citep{yang2025uncertain}.

Introduced by \citet{liu2007uncertainty}, the Liu process, denoted as $C=\{C_t\}_{t\geq 0}$, serves as a fundamental tool within Uncertainty Theory. It can be viewed as the uncertain counterpart of the Brownian motion in probability theory. A Liu process is characterised by several key properties: it starts at zero ($C_0 = 0$), possesses stationary and independent increments, and its sample paths are almost surely Lipschitz continuous. 

Unlike Brownian motion, for any given time increment, the change in the process $C_{s+t} - C_s$ follows a \textit{normal uncertain distribution} with an expected value of $0$ and a variance of $t^2$, with the uncertainty distribution function given by
\begin{equation}
    \label{eq:liuprocess}
    \Phi(x) = \left(1 + \exp\left(-\frac{\pi x}{\sqrt{3}(t - s)}\right)\right)^{-1}.
\end{equation}

This distribution quantifies the belief degree $  \mathcal{M}\{C(t) - C(s) \leq x\} = \Phi(x)  $, where $  \mathcal{M}  $ is the uncertainty measure.

The increments of $C$ are stationary and independent in the uncertainty sense, with almost all sample paths having finite variation, unlike the infinite variation of Brownian motion paths. Compared to the Brownian motion, which relies on probability theory for objective randomness, the Liu process offers advantages in modelling subjective uncertainty through uncertainty theory. It excels in scenarios with limited historical data, allowing direct incorporation of expert beliefs (e.g., in emerging markets or novel risks) rather than requiring frequency-based probabilities. Additionally, its paths have finite variation, making it computationally simpler for simulations and better suited for systems with smoother, less erratic dynamics, such as macroeconomic policy impacts or engineering reliability assessments, while avoiding the Brownian motion's assumption of infinitely many small shocks.

Building upon this process, an Uncertain Differential Equation (UDE) is employed to describe the dynamic evolution of a state variable $S_t$ (such as the number of adopters) over time. A standard UDE takes the following canonical form:
\begin{equation}
    \label{eq:UDE}
    dS_t = f(t, S_t)dt + g(t, S_t)dC_t
\end{equation}
where $f(t, S_t)$ represents the drift term, capturing the deterministic trend of the system, and $g(t, S_t)$ denotes the diffusion term, which scales the impact of the uncertain disturbance $dC_t$.

By substituting the deterministic Bass model's growth rate into the drift term $f(t, S_t)$, \citet{yang2025uncertain} constructed an Uncertain Bass Model. This UDE-based framework comes with robust theoretical analysis, including proofs for the existence and uniqueness of solutions. 

The UBM is justified in contexts where classical SDEs yield counter-intuitive results due to the unbounded variance of the Wiener process. Empirical applications, such as modelling the output of China's integrated circuit industry, suggest UDEs can provide a superior fit compared to traditional stochastic models \citep{li2025modeling}.

\section{Estimation of Bass model parameters}
\label{sec:bass_estimation}

Since its inception, the \citet{bass1969new} model has evolved into an essential framework for characterising the diffusion dynamics of innovations. Its utility lies in providing a robust theoretical basis for market strategists to quantify market potential ($m$) and delineate the interplay between innovators ($p$) and imitators ($q$). Given that the predictive fidelity of the model hinges entirely on these parameters, and as \citet{mitra2019forecasting} mentioned, there is currently a lack of timely estimates of the Bass model parameters that marketers can refer to. The evolution of estimation techniques has remained a focal point of marketing research. Over the past decades, the methodologies for parameterising the Bass model have transitioned from foundational statistical inferences to sophisticated computational heuristics. This section provides a comprehensive review of the estimation techniques employed so far, categorised into three distinct frameworks: 

\begin{enumerate} \item \textbf{Frequentist statistical frameworks:} These encompass the foundational methods used in early Bass model research, primarily focusing on minimising residuals or maximising likelihood functions, such as ordinary least squares and maximum likelihood estimation. Detailed discussions on these approaches are provided in Subsection~\ref{subsec:stat_methods}.

\item \textbf{Bayesian inference and state-space filtering:} This category explores methods that incorporate prior knowledge or dynamic updates to handle parameter uncertainty and time-varying signals, including Bayesian estimation and Kalman filtering techniques. These are elaborated in Subsection~\ref{subsec:bayesian_filtering}.

\item \textbf{Heuristic and meta-heuristic optimisation :} To address the non-convexity and complexity of the parameter space in certain real-world scenarios, modern computational techniques such as genetic algorithms and simulated annealing have been increasingly adopted, due to their their global search capabilities and improved performance. These are discussed in Subsection~\ref{subsec:heuristic_algos}.
\end{enumerate}

Each method has its unique advantages and applicability. By analysing in detail their principles, applications, and performance on actual market data, we aim to provide market researchers with a comprehensive guide to help them choose the parameter estimation method that best suits their needs.

\subsection{Statistical methods}
\label{subsec:stat_methods}

The most established category of parameter estimation for the Bass model relies on frequentist statistical inference, encompassing Ordinary Least Squares (OLS), Nonlinear Least Squares (NLS), and Maximum Likelihood Estimation (MLE). 

\subsubsection{Ordinary least squares}

Historically, the OLS method was the first technique applied to the model \citep{bass1969new}, with its widespread use later reviewed extensively by \citet{mahajan1990new}. Despite the inherent nonlinearity of the Bass model, OLS remains a common tool for preliminary analysis due to its computational simplicity. By rearranging the fundamental model, the adoption at a specific time point can be expressed as a quadratic function of the cumulative adoption:
\begin{equation}
\label{eq:ols}
s(t+1) = \alpha_1 + \alpha_2 S(t) + \alpha_3 S^2(t)
\end{equation}

In this regression framework, $s(t+1)$ denotes the number of adopters (sales) at time $t+1$, while $S(t)$ represents the cumulative number of adopters up to time $t$. The regression coefficients ($\alpha_1, \alpha_2, \alpha_3$) serve as intermediate parameters that map directly to the original Bass coefficients as follows: $\alpha_1 = pm$, $\alpha_2 = q - p$, $\alpha_3 = -q/m$. Once the values of $\alpha_1, \alpha_2$, and $\alpha_3$ are estimated through standard regression analysis, the structural parameters $p$, $q$, and $m$ can be algebraically recovered. 

This approach facilitates rapid estimation, particularly when initial market data is sparse \citep{srinivasan1986nonlinear}, and provides a baseline for exploring factors such as time-series length and market potential estimates \citep{jiang2006virtual}. However, the simplicity of OLS comes with significant trade-offs. \citet{horsky1983advertising} pinpointed critical issues such as endogeneity bias and the tendency of OLS to underestimate the imitation effect ($q$) compared to nonlinear alternatives. 

\subsubsection{Nonlinear least squares}

To rectify these biases, the Nonlinear Least Squares (NLS) method has become the preferred standard for many researchers. NLS directly addresses the nonlinear structure of the cumulative adoption function by minimizing the sum of squared differences between observed and predicted values:
\begin{equation}
\label{eq:nls}
\min_{p, q, m} \sum_{t=1}^T \left( S(t) - \hat{S}(t; p, q, m) \right)^2
\end{equation}

Equation~\eqref{eq:nls} defines the objective function for NLS estimation, which aims to minimise the aggregate squared discrepancy between observed cumulative adoption $S(t)$ and the theoretical trajectory $\hat{S}(t; p, q, m)$ suggested by the model. By seeking the values of $p$, $q$, and $m$ that yield the smallest sum of squared residuals, this approach ensures that the resulting diffusion curve provides the best numerical fit to empirical market data across the entire observation horizon $T$. Unlike linearised methods, NLS directly preserves the structural nonlinearities of the Bass model, thereby capturing the S-shaped innovation diffusion process with higher fidelity.

Early work by \citet{easingwood1983nonlinear} and \citet{srinivasan1986nonlinear} demonstrated that NLS is more robust than OLS in capturing the diffusion dynamics of both industrial and consumer innovations. While \citet{little1970models} focused more broadly on decision calculus, his work laid the conceptual groundwork for adopting such sophisticated nonlinear methods to enhance marketing decision-making. 

Later comparative studies by \citet{kumar1995comparative} further validated the robustness of NLS in handling the complexities of real-world adoption data. The versatility of NLS is evidenced by its extensive application across diverse industries. In the pharmaceutical and healthcare sectors, NLS has been used to model the adoption of medical devices and competitive interactions between drugs \citep{guseo2009market, guseo2015modelling}. 

Similarly, NLS has proved effective in analysing information technology services \citep{talukdar2002product}, mobile telecommunications—where network effects and competition are paramount \citep{jiang2010dynamics}—and educational technology, helping providers optimise promotion strategies based on institutional adoption drivers \citep{roberts2005generalizing}.

\subsubsection{Maximum likelihood estimation}

Moving beyond residual minimisation, Maximum Likelihood Estimation (MLE) offers a rigorous statistical framework that maximises the probability of observing the given data under the assumed model. 

First introduced to the Bass model by \citet{schmittlein1982maximum}, MLE treats the adoption process as a probabilistic sequence. The likelihood function is defined by the joint probability density of all observed adoptions, formulated as:
\begin{equation}
\label{eq:mle}
L(p, q, m) = (1-F(T-1))^{s(T)}\prod_{t=1}^{T} [F(t)-F(t-1)]^{s(t)}
\end{equation}

By maximising the log-likelihood $\log L(p, q, m)$, researchers leverage the property of asymptotic efficiency, which ensures that Maximum Likelihood Estimators (MLE) achieve the lowest possible variance among consistent estimators in large samples. This property often results in narrower confidence intervals and more powerful hypothesis tests compared to least squares methods \citep{greene2000econometric}.

The MLE framework is particularly adept at incorporating external covariates. For instance, it has been used to estimate the impact of advertising \citep{horsky1983advertising}, price effects for durable goods \citep{jain1990effect}, and the biases inherent in macro-level diffusion estimates \citep{vandenbulte1997bias}. More recently, \citet{vandenbulte2004social} utilised MLE to disentangle the effects of social contagion from income heterogeneity. Despite its strengths, the validity of MLE depends on the correctness of the assumed probability distribution. Unlike least squares estimators, which remain consistent under weaker conditions, MLE estimates can be inconsistent if the underlying distributional assumption is misspecified \citep{greene2000econometric}.

\subsection{Bayesian filtering methods}
\label{subsec:bayesian_filtering}

Unlike traditional methods that treat parameters as fixed constants, Bayesian estimation incorporates prior knowledge—such as expert opinions or historical data from similar product categories—into the probability distribution of $p$ and $q$. 

\subsubsection{Bayesian updating}

First introduced to the innovation diffusion field by \citet{mahajan1984estimation}, the Bayesian framework utilises Bayes' theorem to derive a posterior distribution:
\begin{equation}
\label{eq:bayesian}
\pi(p, q \mid \text{data}) \propto L(\text{data} \mid p, q) \pi(p, q)
\end{equation}

where the prior $\pi(p, q)$ (often modelled as Beta or Normal distributions) is updated by the likelihood function $L(\text{data} \mid p, q)$ as new market observations emerge. 

This methodology has proven robust across diverse sectors, including consumer electronics \citep{inman1994coupon}, biotechnology \citep{golder1997will}, and automotive technologies \citep{hauser2006research}. It is particularly valuable in global market analysis, where researchers must integrate cultural and economic heterogeneity across borders \citep{peres2010innovation, venkatesan2004customer}, and in the pharmaceutical industry, where clinical trial data serves as a critical prior for forecasting drug adoption \citep{montoya2010dynamic}. 

\citet{arts2011generalizations} showed that advanced Bayesian techniques such as Markov Chain Monte Carlo (MCMC) enable accurate estimation in the case of environmentally friendly technologies even when early-market data is scarce. More recently, the framework has been extended to simulate propagation paths in social media marketing, identifying key drivers of information diffusion through large-scale network data \citep{libai2013decomposing}.

\subsubsection{Kalman filtering}

While Bayesian updating is designed to update probability distributions, adaptive and Kalman filtering techniques focus on the recursive tracking of time-varying parameters. Adaptive filtering assumes that the coefficients of the discretised Bass model, $\hat{x}_i(t)$, evolve over time to reflect changing market patterns:
\begin{equation}
\label{eq:adaptive_filtering}
\hat{x}_i(t) = \hat{x}_i(t - 1) + X_i(e(t))
\end{equation}

Here, the feedback filter $X_i(e(t))$—typically implemented via the Least Mean Squares algorithm or the Carbone-Longini Adaptive Estimation Process \citep{bretschneider1980adaptive}—adjusts the parameters based on the one-step-ahead forecast error $e(t)$. 

As noted by \citet{bretschneider1980adaptive}, these adaptive models are superior for technology substitution forecasting where parameter shifts are unknown and require automatic adjustment to data patterns.

The Kalman Filter (KF) further refines this recursive logic by treating the Bass parameters and market potential as a dynamic state vector $x_t = (p, q, m)^\top$ at time $t$. KF operates as a recursive estimator that tracks the evolution of the Bass model parameters by alternating between a time-update (prediction) phase and a measurement-update (correction) phase. The core of this mechanism lies in the following equations, which refine the state estimate by integrating the latest market adoption data:
\begin{equation}
\label{eq:kf_update}
\hat{x}_{t|t} = \hat{x}_{t|t-1} + K_t (s_t - H_t \hat{x}_{t|t-1})
\end{equation}
\begin{equation}
\label{eq:kf_weight}
K_t = P_{t|t-1} H_t^\top (H_t P_{t|t-1} H_t^\top + R_t)^{-1}
\end{equation}

In these expressions, $H_t$ is the observation matrix, which defines the functional relationship between the internal state vector (the Bass parameters $p, q$, and $m$) and the observable adoption data $s_t$. $P_{t|t-1}$ denotes the predicted error covariance matrix, representing the estimated uncertainty or variance associated with the state prediction before the current measurement is incorporated. By utilising $H_t$ to calculate the ``innovation'' (the gap between actual and predicted sales) and $P_{t|t-1}$ to derive the optimal Kalman gain $K_t$, the filter effectively balances the reliability of existing model trajectories against the measurement noise $R_t$ found in empirical data.

However, a critical limitation of the standard Kalman Filter is its inherent assumption of linearity in both the state transition and observation equations. The Bass model is fundamentally nonlinear, particularly when the market potential $m$ and the coefficients $p$ and $q$ are treated as time-varying states, as the expected sales $s_t$ depend on the product of these state variables and the cumulative adoption. Applying the standard linear update~\eqref{eq:kf_update} to this nonlinear structure can result in significant estimation errors or filter divergence. To mitigate this, the Extended Kalman Filter (EKF) is typically employed in diffusion literature. The EKF addresses the nonlinearity by linearising the observation function around the current state estimate using a first-order Taylor series expansion.

The comprehensive review \citet{bensaid2011comparison} highlights that the evolution from basic KF to EKF and Recursive Prediction Error methods allows for a sophisticated handling of the inherent nonlinearities and noise in innovation diffusion. Moreover, \citet{xie1997kalman} demonstrates that the Augmented Kalman Filter (AKF) provides significantly higher accuracy for period-by-period adoption forecasting. This recursive power has been validated in broader marketing literature for production curve estimation \citep{meade1985forecasting} and
sales rate prediction \citep{munroe2009sales}.

\subsection{Heuristic algorithms}
\label{subsec:heuristic_algos}

To overcome the limitations of traditional optimisation methods, such as the propensity to become trapped in local minima, researchers have increasingly turned to meta-heuristic frameworks. 

\subsubsection{Genetic algorithms}

Genetic Algorithms (GA) represent a prominent bio-inspired approach that simulates the principles of natural selection, crossover, and mutation to explore the parameter space. In the context of the Bass model, the algorithm evaluates the ``fitness'' of various combinations of $(p, q, m)$—defined by their alignment with empirical market data—and iteratively evolves these populations to identify the global optimum.

The efficacy of GA-based estimation has been validated across several complex empirical settings. \citet{tsai2014forecasting} utilised a hybrid GA-NLS approach to analyse industrial clusters in China, demonstrating that the integrated model provides significantly higher stability and precision than GA alone after extensive simulation. Furthermore, \citet{venkatesan2002genetic} found that while gradient-based methods are efficient for smooth solution surfaces, GA-NLS is superior in avoiding local optima when predicting mobile phone penetration across Europe. 

Crucially, research by \citet{venkatesan2004evolutionary} suggests that GA-NLS outperforms Bayesian and Kalman filtering techniques when only pre-peak sales data are available, providing unbiased estimates where other models might fail.

\subsubsection{Simulated annealing}

Complementing these population-based methods is Simulated Annealing (SA), a trajectory-based heuristic that simulates the thermodynamic process of metal cooling to find a system's minimum energy state. The algorithm explores the neighbourhood of a current solution and identifies a change in the cost function, $\Delta L$. In the context of the Bass model, this cost function is typically the sum of squared residuals:
\begin{equation}
\label{eq:sa}
L = \sum_{t=1}^{n} (\hat{s}_t - s_t)^2
\end{equation}

where $s_t$ is the actual observation of noncumulative number of adopters during $(t,t+1)$ and $\hat{s}_t$ is the theoretical value.

A unique feature of SA is its probabilistic acceptance of inferior solutions—specifically with a probability defined by $e^{-\Delta L / T}$—allowing the search to ``jump'' out of local minima during the high-temperature phase. 

As the ``temperature'' $T$ gradually decreases, the search stabilises toward the global optimum. This technique is particularly effective for highly nonlinear variants of the Bass model, such as the Parameter Variability Randomness in Diffusion (PVRD) model explored by \citet{goswami2004study}, for which the parameter landscape can be very complex. 

Ultimately, the choice of parameter estimation technique depends on the specific characteristics of the market data and the structural complexity of the model. While classical frequentist methods (OLS, NLS, MLE) provide a foundation for rapid analysis, Bayesian and filtering frameworks offer the flexibility to incorporate prior knowledge and time-varying signals. In scenarios defined by high nonlinearity or limited early-stage data, heuristic optimisations such as GA and SA ensure the robustness and global validity of the diffusion forecast. By integrating these diverse methodological strengths, researchers can achieve a higher degree of predictive fidelity in modelling the lifecycle of new product innovations.

\section{Other innovation diffusion models}
\label{sec:others}

\subsection{Substitution models}

Innovation diffusion can also be described as an evolutionary process, where an old technology is replaced by a new one to solve similar problems or achieve similar goals. The theory of substitution is based on the competition between old and new technologies, or simply put, it can be seen as the process of adopting a new innovation or product to replace existing ones.

The process of ``new replacing old'' is not unfamiliar. Apart from the Bass model, many other models have been proposed over the years to describe and represent the substitution process over time, or in other words, the innovation diffusion process. Among the most fundamental deterministic models are the Logistic, Gompertz, and Weibull distributions. 

The Logistic model \citep{verhulst1838notice}, originally derived to describe population growth under resource constraints, posits that the rate of adoption is driven by the interaction between current adopters and potential adopters (a ``contagion'' effect). Its formulation is given by:
\begin{equation}
    \label{eq:logistic}
    F(t) = \frac{1}{1 + e^{-(a + bt)}}, \quad S(t) = \frac{m}{1 + e^{-(a + bt)}}
\end{equation}
where $m$ represents the potential market size (saturation level). The parameter $a$ is a location parameter that determines the horizontal shift of the curve (related to the timing of the inflection point), while $b$ serves as the growth rate parameter, controlling the steepness or speed of the diffusion process.

In contrast, the Gompertz model \citep{gompertz1833nature}, initially developed to analyze human mortality rates, implies that the adoption rate decreases exponentially as the market approaches saturation, making it suitable for technologies that diffuse rapidly at first but face resistance or ``friction'' in the late stages. The model is defined as:
\begin{equation}
    \label{eq:gompertz}
    F(t) = e^{-e^{-(a + bt)}},\quad S(t) = m \cdot e^{-e^{-(a + bt)}}
\end{equation}
Here, $b$ dictates the rate of growth decay, and $a$ sets the initial displacement. Compared to the Logistic curve, the inflection point of the Gompertz curve occurs earlier, reflecting a front-loaded adoption pattern.

Additionally, the Weibull model \citep{weibull1951statistical} draws from reliability engineering and failure analysis. It treats the adoption of a new product as a ``time-to-failure'' event for a non-adopter. This model offers significant flexibility through its shape parameter, allowing it to accommodate various diffusion patterns:
\begin{equation}
    \label{eq:weibull}
    F(t) = 1 - e^{-(\frac{t}{a})^b},\quad S(t) = m \left[ 1 - e^{-(\frac{t}{a})^b} \right]
\end{equation}
In this function, $a$ is the scale parameter, influencing the overall timing of the diffusion, while $b$ is the shape parameter. When $b > 1$, the curve is S-shaped; when $b = 1$, it simplifies to an exponential distribution; and when $b < 1$, it represents a concave diffusion curve.

Finally, the Shifted Gompertz model \citep{bemmaor1994modeling} introduces the perspective of consumer heterogeneity regarding the propensity to buy. Unlike aggregate-level models, it is built upon individual-level assumptions where the propensity to adopt follows a specific probability density function:
\begin{equation}
    \label{eq:shifted_gompertz}
    f(t) = b e^{-bt} \exp( -\eta e^{-bt}) [1 + \eta (1-e^{-bt})]
\end{equation}
where $b$ is a scale parameter ($b>0$) and $\eta$ is a shape parameter ($\eta>0$). This model postulates that early adopters purchase in a more random fashion, whereas late adopters exhibit more deterministic behavior. As the shape parameter $\eta$ approaches 0, the density converges toward an exponential distribution, providing a nuanced view of the underlying purchasing mechanism.

Following the establishment of these foundational theories, comparative studies emerged. \citet{winsor1932gompertz} compared the Gompertz process and the logistic process as a growth curve. \citet{kumar1995comparative} and \citet{meade2006modelling} both surveyed a large number of such models. Some of these models draw inspiration from traditional biological processes (such as species competition, cell or disease diffusion), while others are influenced by classical mathematical processes (such as S-shaped curves).

From a biological perspective, this replacement process can be mathematically defined by the Lotka-Volterra predator-prey equations  (\citealt{lotka1925elements}, \citealt{volterra1926fluctuations}). From a medical standpoint, the Gompertz model has been widely used to simulate the diffusion of disease cells in the human body. Both models have been extensively applied to innovation diffusion, see for example \citet{gutierrez2007new},
\citet{tjorve2017gompertz}, and \citet{utterback2018dynamics}. Moreover, pandemic models such as SIR models (Susceptible-Infected-Removed) can also be seen as models for gradual replacement. As an example, \citet{mandl2023innovations} compares them in several perspectives.

From the perspective of technological diffusion, the development of substitution models originated with \citet{mansfield1961technical} and \citet{fisher1971simple}. The Bass model is also well-known as a prominent branch of this framework. \citet{mansfield1961technical} made pioneering contributions to the development of substitution models, establishing a model to describe the substitution process in industrial sectors. \citet{floyd1962trend} and \citet{sharif1976system} further expanded these models, representing general models for improving technological substitution forecasts. It is worth noting that these models consistently adopt the logistic equation form because the S-shaped form exhibits many characteristics consistent with technological diffusion in reality, such as the rapid acceleration followed by a slowdown in diffusion speed and a diffusion ceiling. \citet{bewley1988flexible} designed the Flexible-Logistic Model (Flog), further expanding the application of the logistic equation in innovation diffusion. Moreover, \citet{sharif1980weibull} drew inspiration from the Weibull distribution and applied it to the study of innovation diffusion. This distribution was originally used to describe the intervals between random events (such as rainfall or failures).

\subsection{Agent-based approach}
\label{subsec:agent_based}

\subsubsection{Bottom-up and top-down approaches}
\label{subsubsec:bottom_up_top_down}

The models mentioned earlier, whether deterministic Bass models, extended models, or their stochastic versions, all employ a Top-Down Approach. The core idea of this approach is to focus on overall changes and describe how innovation (in terms of adopter proportions, etc.) spreads from a systems dynamics perspective. This approach brings great convenience both from a simulation and parameter estimation standpoint. However, it also faces various challenges in application. For example, when potential adopters have unpredictable decision logic or when there is strong interaction among participants in the market, this macroscopic approach can fail to yield the desired predictive results. An alternative modelling approach, known as agent-based modelling (ABM), adopts a different strategy to address such issues. 

\begin{figure}[ht]
    \centering
    \includegraphics[scale=0.6]{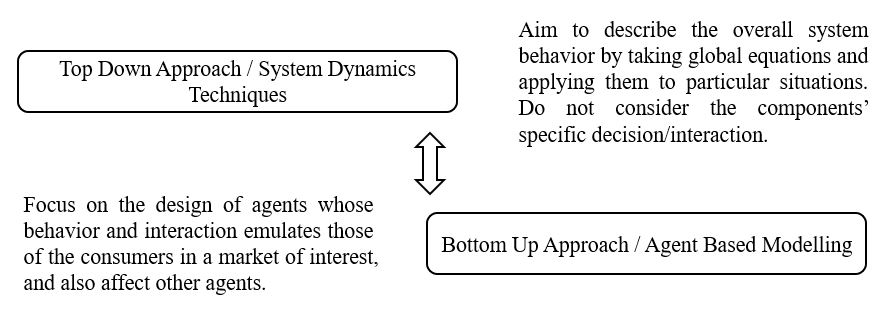}
    \caption{Difference between the two strategies}
    \label{fig:top_down_vs_bottom_up}
\end{figure}

\cite{asgharpour2010impact} reviewed the impact and the widespread use of ABM in social science. This strategy focuses on simulating individual decision logic and interactions among individuals, simulating the performance of different individuals after the introduction of a new product, and finally observing the behaviour of the system in simulation results. Therefore, this approach is also referred to as the Bottom-Up Approach. Figure \ref{fig:top_down_vs_bottom_up} illustrates the differences between the two approaches.

Recently, agent-based modelling has gained traction for its ability to capture complex emergent phenomena at the individual level, particularly in the diffusion of innovative products, where numerous contributions have been made in the literature. \citet{dawid2006agent} provides a review of specific agent-based computational economics models in innovation and technological revolutions. \citet{garcia2005uses} reviews the applications of ABM in the field of innovation diffusion, outlining potential applications and highlighting several major challenges faced by ABM, including the difficulty in designing interactions between agents, the quantitative prediction shortcomings of the models themselves, and the high computational costs of complex models. \citet{kiesling2012agent} offers a systematic review of past ABM literature, categorising it based on the decision rules adopted and interaction models. \citet{holanda2003modeling} focuses on the Bass model, providing a detailed comparison of two different strategies and their predictive results.

\subsubsection{Agent-based modelling for innovation diffusion}
\label{subsubsec:agent_based}

In this part, we aim to review and categorise some past cases of applying ABM to innovation diffusion. Two crucial aspects in the specific implementation of ABM are decision rules and interaction patterns among agents. Decision rules delineate the reasons and thresholds for individuals to adopt new products, while interaction patterns describe how the opinions and decisions of each individual affect others. From these two perspectives, ABM in innovation diffusion can be classified into the following categories:

\begin{enumerate}
    \item Decision rules:

    \begin{itemize}
        \item \textbf{Decide based on imitation}: This is the simplest and most common decision rule, consistent with the assumptions of the Bass model. Imitators make adoption decisions based on the behaviour of other agents. Typically, a threshold is designed, and once the proportion of adopters in the market reaches a certain level, individuals will make (or have a probability of making) imitation decisions.

        \item \textbf{Decide based on utility}: A more rational perspective where agents not only imitate but also consider the impact of adopting the new product. Specifically, individuals have an expected utility, and when the benefits of adopting the new product (product quality, price) meet expectations, individuals choose to adopt the new product.

        \item \textbf{Transition between multiple states}: Relatively more complex and closer to the original version of innovation diffusion theory, where agents possess not only two states (``potential adopter'' and ``adopter'') but multiple states (such as awareness, information seeking, and adopter) and transition between them based on environmental influences.

        \item \textbf{Social psychology approach}: Modelling consumer agents' behaviour based on Ajzen's Theory of Planned Behaviour (TPB,  \citealt{ajzen1991theory}). It postulates that attitude, perceived behavioural control, and intention are predictors of behaviour.
    \end{itemize}

    \item Interaction level:

    \begin{itemize}
        \item \textbf{Local knowledge level}: It assumes that potential users are influenced by advertising and by their close neighbours but may not know the proportion of adopters globally. At this level, word of mouth plays a major role.

        \item \textbf{Global knowledge levels}: It assumes that agents are influenced not only by neighbours but also by some global information, such as the influence of aggregate network-level opinion and macroeconomic feedbacks.
    \end{itemize}
\end{enumerate}

The Imitation Decision Rule is the most widely adopted decision-making approach. For instance, in \citet{valente1999accelerating}, individuals are assumed to imitate if over 15\% of their personal network adopts a particular innovation. Similarly, \citet{decanio2000importance} and \citet{alkemade2005strategies} make similar assumptions, albeit with differences in threshold design. The core idea here is that individuals' decisions are influenced by the decisions of those around them. \citet{bohlmann2010effects} makes slight adjustments, positing that agents adopt with a certain probability once the threshold is reached, still aligning with this type of decision rule.

The Utilitarian Decision Rule differs in that economic benefits and utility factors are key drivers of individual decision-making in reality. This type of decision rule may be more applicable to businesses adopting new technologies (such as communication systems or production lines). \citet{delre2007targeting} provides a utility function where consumers opt for adoption when a new product exceeds their quality expectations. \citet{keeney1993decisions} makes a simple attempt to integrate multi-attribute preference modelling approaches with ABMs of innovation diffusion.

Unlike models describing agents with only two states, \citet{goldenberg2001using} and \citet{deffuant2005individual} depict adopters in multiple states, influenced by word-of-mouth factors, leading agents to transition between different states. \citet{thiriot2008using} breaks down the process of agent adoption of new products into more stages. The Social Psychology approach also differs from previous models, drawing on \citet{ajzen1991theory}'s Theory of Planned Behaviour (TPB) as one of its foundations. It assesses agents' attitudes and perceptions to determine decision-making. For example, \citet{kaufmann2009simulating} utilises TPB to model the diffusion of organic farming practices, in which farmers were assumed to adopt new technology if their intention exceeded an empirically derived threshold. \citet{schwarz2009agent} applies ABM to model consumers' decisions to adopt or reject water-saving innovations using two different decision rules based on TPB.

Regarding interactions among agents, while they can be categorised into global and local levels, many articles do not strictly adhere to one or the other; instead, they blend aspects of both. For example, an agent may be influenced by word-of-mouth effects from neighbours while also being swayed by advertising and social evaluations when deciding to adopt a new product. In terms of specific design, concepts such as ``Moore neighbours'' from \citet{gilbert2005simulation} and the ``small world'' from \citet{watts2004small} have provided significant assistance in model development and are critical theoretical underpinnings supporting the models.

Recent ABM research has challenged the macro-model assumption of consumer homogeneity and simple network structure. Analysis using the stochastic discrete Bass model reveals that while heterogeneity in only the innovation ($p_j$) or imitation ($q_j$) coefficient generally slows diffusion, heterogeneity in both $p_j$ and $q_j$ (provided they are not positively correlated) can, surprisingly, accelerate the aggregate diffusion rate \citep{fibich2022diffusion}.
Leveraging computational techniques (e.g., Python/NetLogo integration), ABM allows for diffusion simulation on complex, realistic networks \citep{dilucchio2024bass}. Studies show that assortative networks (where high-degree nodes connect to each other)—a common feature of social systems—tend to delay the adoption peak time compared to uncorrelated networks. Furthermore, extending the model to signed networks that incorporate negative influence (or ``foe relationships'') via a threshold rule \citep{mueller2023signed} shows that diffusion is significantly hindered, resulting in adoption blockades. This negative effect is exacerbated in networks exhibiting high clustering.

\section{Conclusion}
We are now living in a world of perpetual innovation, with the advent of technologies such as cloud computing, 5G connectivity, the internet of things, blockchain, generative artificial intelligence, and many other emerging technologies. The study and management of innovation risks are becoming ever more important, yet financial valuation models that rely solely on historical financial information and traditional asset pricing theory are increasingly becoming inadequate, as they fail to account for the complex nature of expected financial rewards from potential innovations and their underlying risks.

As an important step forward, this paper seeks to address this issue by examining a particular angle on innovation risks, namely the risks of innovation diffusion. History has witnessed many successes and failures of innovations, which have inspired ongoing research on innovation diffusion, as epitomised by E.M. Rogers' theory of innovation diffusion \citep{rogers1962innovation}. The seminal work of \citet{bass1969new} and its later variants and extensions provide a quantitative framework for modelling innovation diffusion risks that embodies the key insights from Rogers' theory. 

We have conducted an extensive meta-study that synthesises existing research on the dynamics of innovation diffusion using Bass-type models, with the ultimate goal of understanding, quantifying and managing innovation risks under a reliable quantitative framework. We described the seminal work of \citet{bass1969new} and its various structural and stochastic extensions, as well as its parameter estimation techniques. We also briefly explored non-Bass innovation diffusion models, with a particular focus on agent-based models as a \textit{bottom-up} counterpart to the aforementioned \textit{top-down} Bass-type models.

Potential limitations of the models described in this review can stem from data issues, such as missing, censored, or inaccurate data, from invalid simplifying assumptions, for example time-varying preferences or heterogeneities between different sectors or countries, or from external economic shocks. More general stochastic models such as jump-diffusion models and their integration within, for example, a real options framework could be investigated in the context of innovation diffusion and IP valuation. These open questions are left for future research.

\section*{Acknowledgements}

Our work was supported by the Guangdong Provincial/Zhuhai Key Laboratory of IRADS (2022B1212010006).

\bibliographystyle{plainnat_lastnamefirst.bst}
\bibliography{bibliography}

\end{document}